\documentclass[twocolumn,prb,preprintnumbers,amsmath,amssymb,letterpaper,floatfix,showpacs]{revtex4-1}
\usepackage{graphicx} 
\usepackage{times}
\usepackage{gensymb } 
\usepackage{graphicx} 
\usepackage{subfigure} 
\usepackage{ upgreek } 
\usepackage{float} 
\usepackage{natbib}

\begin{document}
\title{Pressure-Induced Phase Transition in La$_{1-x}$Sm$_{x}$O$_{0.5}$F$_{0.5}$BiS$_2$}

\author{Y. Fang$^{1,2}$}
\author{D. Yazici$^{2,3}$}
\author{B. D. White$^{2,3}$}
\author{M. B. Maple$^{1,2,3}$}
\email[Corresponding Author: ]{mbmaple@ucsd.edu}

\affiliation{$^1$Materials Science and Engineering Program, University of California, San Diego, La Jolla, California 92093, USA}
\affiliation{$^2$Center for Advanced Nanoscience, University of California, San Diego, La Jolla, California 92093, USA}
\affiliation{$^3$Department of Physics, University of California, San Diego, La Jolla, California 92093, USA}
\date{\today}

\begin{abstract}
Electrical resistivity measurements on La$_{1-x}$Sm$_{x}$O$_{0.5}$F$_{0.5}$BiS$_{2}$ ($x$ = 0.1, 0.3, 0.6, 0.8) have been performed under applied pressures up to 2.6 GPa from 2 K to room temperature. The superconducting transition temperature \textit{T$_{c}$} of each sample significantly increases at a Sm-concentration dependent pressure \textit{P$_{\rm t}$}, indicating a pressure-induced phase transition from a low-\textit{T$_{c}$} to a high-\textit{T$_{c}$} phase. The compounds that have higher Sm concentrations have higher \textit{T$_{c}$} values at ambient pressure; however, the \textit{T$_{c}$} values at $P$ $\textgreater$ \textit{P$_{\rm t}$} decrease with $x$ and \textit{P$_{\rm t}$} shifts to higher pressures with Sm substitution. In the normal state, semiconducting-like behavior is suppressed and metallic conduction is induced with increasing pressure in all of the samples. These results suggest that the pressure dependence of \textit{T$_{c}$} for the BiS$_{2}$-based superconductors is related to the lattice parameters at ambient pressure and enable us to estimate the evolution of \textit{T$_{c}$} for SmO$_{0.5}$F$_{0.5}$BiS$_{2}$ under pressure.
\end{abstract}
\pacs{61.50.Ks, 74.25.F-, 74.62.Fj, 74.70.Dd}

\maketitle


\section{INTRODUCTION}
The application of external pressure to materials has led to the discovery of many new superconductors and has apparently raised the record of the highest superconducting transition temperature, \textit{T$_{c}$}, to 190 K in H$_{2}$S.\cite{drozdov2014conventional} Due to its substantial impact on the crystalline and electronic structure of solids, applied pressure is recognized as a powerful tool to tune the \textit{T$_{c}$}, critical fields, and other physical properties of superconductors.\cite{shimizu,sefat} Recently, superconductivity with \textit{T$_{c}$} values ranging from 2.7 to 10.6 K at ambient pressure has been reported for BiS$_{2}$-based compounds including Bi$_{4}$O$_{4}$S$_{3}$, \textit{Ln}O$_{1-x}$F$_{x}$BiS$_{2}$ (\textit{Ln} = La, Ce, Pr, Nd, Yb), La$_{1-x}$\textit{M}$_{x}$OBiS$_{2}$ (\textit{M} = Ti, Zr, Hf, Th), and Sr$_{1-x}$La$_{x}$FBiS$_{2}$.\cite{Mizuguchi1,Singh,Awana1,Demura1,Mizuguchiz&Yoshikau,Xing,Yazici1,Mizuguchi&Yoshikazu2,Deguchi,Yazici2,Lin} Similar to the cuprate and Fe-based superconductors, the structure of these compounds is characterized by alternate stacking of superconducting BiS$_{2}$ layers and charge-reservoir blocking layers, both of which are tunable by using chemical substitution or applying external pressure.\cite{Mizuguchi1,Singh,Mizuguchi&Yoshikazu2,Lin,yazici2015superconductivity,mizuguchi2014review} Therefore, there is plenty of phase space to search for the optimal conditions for superconductivity in this family of compounds.

The compound LaO$_{0.5}$F$_{0.5}$BiS$_{2}$, synthesized under high pressure, was reported to have the highest \textit{T$_{c}$} of $\sim$10.6 K in contrast to the \textit{T$_{c}$} $\sim$ 3 K of the same compound when it is synthesized at ambient pressure.\cite{mizuguchi2014stabilization,Deguchi} Although the crystal structure of both the sample annealed at ambient pressure (AP) and the one synthesized under applied high pressure (HP) have the same space group \textit{P4/nmm}, the strongly reduced lattice parameters $a$ and $c$ are considered to be related to the enhancement in \textit{T$_{c}$} of HP samples.\cite{mizuguchi2014stabilization,Deguchi,PallecchiI} On the other hand, LaO$_{0.5}$F$_{0.5}$BiS$_{2}$ (AP), which has a low \textit{T$_{c}$}, undergoes a structural phase transition from tetragonal to monoclinic at a pressure of 0.7 GPa, and \textit{T$_{c}$} is enhanced to 10.7 K.\cite{TakahiroTomita} Similar abrupt increases in \textit{T$_{c}$} were also observed in the compounds \textit{Ln}O$_{0.5}$F$_{0.5}$BiS$_{2}$ (\textit{Ln} = Ce, Pr, Nd) and Eu$_{3}$Bi$_{2}$S$_{4}$F$_{4}$ under pressure.\cite{Wolowiec1,Wolowiec2,luo2014pressure} A gradual increase in \textit{T$_{c}$} up to 5.4 K was observed at ambient pressure in La$_{1-x}$Sm$_{x}$O$_{0.5}$F$_{0.5}$BiS$_{2}$ with increasing Sm concentration until the solubility limit near $x$ = 0.8.\cite{Fang} The lattice parameter $a$ in La$_{0.2}$Sm$_{0.8}$O$_{0.5}$F$_{0.5}$BiS$_{2}$ is significantly smaller compared with other BiS$_{2}$ based compounds; however, further optimization of superconductivity in this system is prevented by the presence of a solubility limit.\cite{Fang}  Hence, it would be interesting to explore the electrical transport properties of the system La$_{1-x}$Sm$_{x}$O$_{0.5}$F$_{0.5}$BiS$_{2}$ under pressure. Such an experiment will not only provide a broad picture of how \textit{T$_{c}$} evolves at extreme conditions, but will also help to determine which parameters are essential to promote superconductivity in the BiS$_{2}$-based compounds. Additionally, pressure may be particularly well suited to induce certain types of phase transitions in La$_{0.2}$Sm$_{0.8}$O$_{0.5}$F$_{0.5}$BiS$_{2}$ due to its extraordinarily small lattice parameters in ambient conditions.

\section{EXPERIMENTAL DETAILS}
The synthesis and crystal structure analysis of polycrystalline samples of La$_{1-x}$Sm$_{x}$O$_{0.5}$F$_{0.5}$BiS$_{2}$ are described elsewhere.\cite{Yazici1,Fang} The chemical composition of each sample presented in this paper is nominal; impurity phases, including La(Sm)F$_{3}$, La(Sm)O, and Bi$_{2}$S$_{3}$, typically $\sim$4 wt.$\%$ for $x$ = 0.1, 0.3, 0.6 and $\sim$8 wt.$\%$ for $x$ = 0.8, were found in these samples.\cite{Fang} Geometric factors for each sample were measured before applying pressure and used to calculate the electrical resistivity from measurements of resistance. Electrical resistance measurements from ambient pressure to $\sim$2.6 GPa were performed on samples with nominal Sm concentrations $x$ = 0.1, 0.3, 0.6, 0.8 between 2 and 280 K using a standard four-probe method in a pumped $^{4}$He dewar. Hydrostatic pressures were generated by using a clamped piston-cylinder cell in which a 1:1 by volume mixture of n-pentane and isoamyl alcohol was used as the pressure transmitting medium. The pressures applied to the samples were inferred from the \textit{T$_{c}$} of a high purity (\textgreater 99.99 \%) Sn disk inside the sample chamber of the cell using the well-established behavior of \textit{T$_{c}$($P$)} in high purity Sn.\cite{Smith}

\begin{figure*}[t]
\centering
\includegraphics[width=15cm]{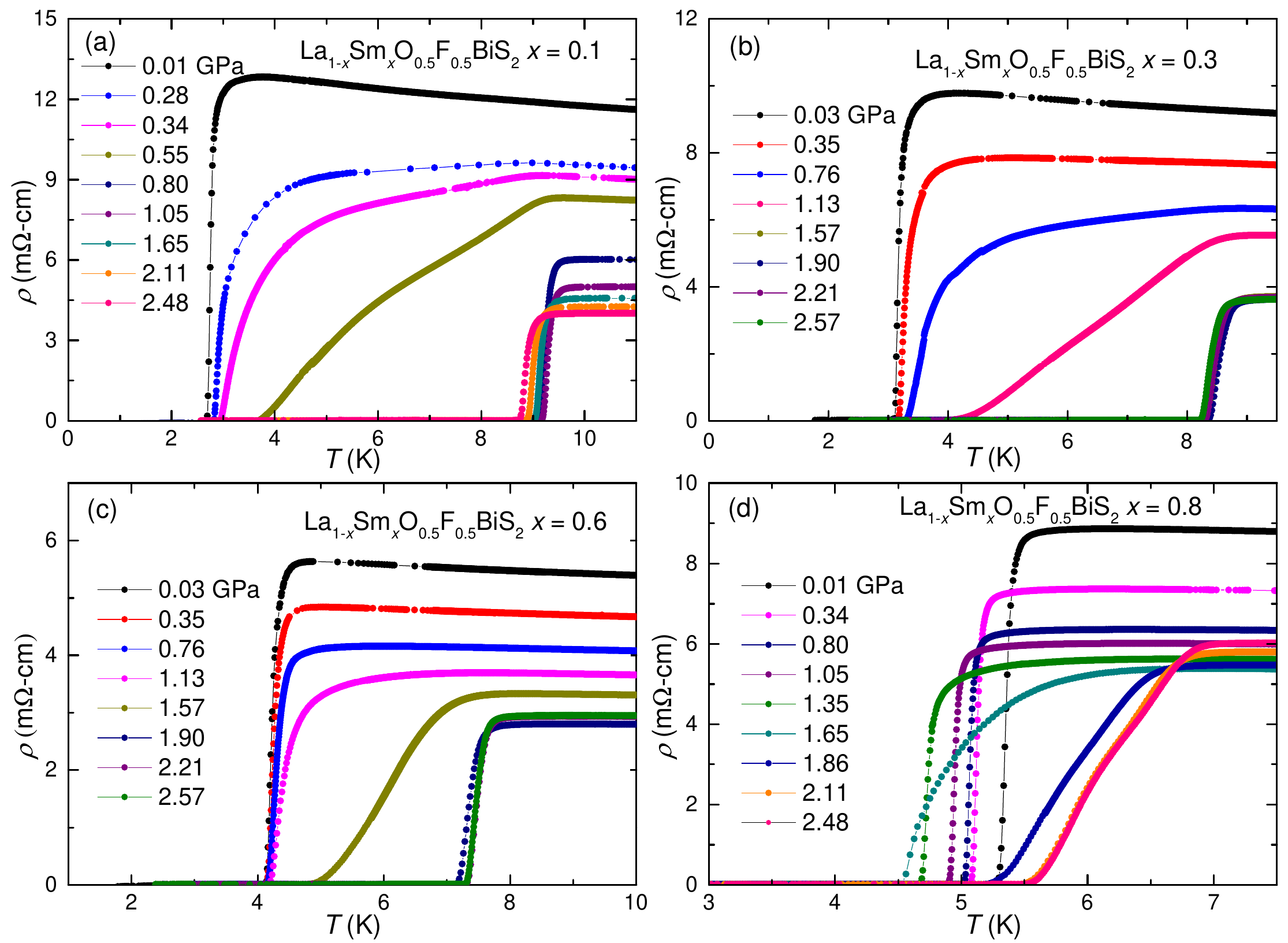}
\caption{(Color online) (a), (b), (c), (d) Temperature dependence of the electrical resistivity at various applied pressures near \textit{T$_{c}$} for La$_{1-x}$Sm$_{x}$O$_{0.5}$F$_{0.5}$BiS$_{2}$ with $x$ = 0.1, 0.3, 0.6, and 0.8, respectively. Lines are guides to the eye.}
\label{FIG.1.}
\end{figure*}

\section{RESULTS AND DISCUSSION}

Measurements of the low-temperature electrical resistivity $\rho$($T$) for La$_{1-x}$Sm$_{x}$O$_{0.5}$F$_{0.5}$BiS$_{2}$ ($x$ = 0.1, 0.3, 0.6, 0.8) samples in zero magnetic field under applied pressures are depicted in Fig. 1. Superconducting (SC) transitions in which $\rho$ abruptly drops from a finite value to zero were observed in each sample from ambient pressure to the highest pressure ($\sim$2.6 GPa) applied in this study. We defined \textit{T$_{c}$} as the temperature where the electrical resistivity ($\rho$) falls to 50\% of its normal-state value, and the width of the transition is characterized by identifying the temperatures where the electrical resistivity decreases to 90\% and 10\% of its normal state value. Superconducting transitions are very sharp at low pressures for all of the samples. However, as can be seen in Fig. 1, the width of the SC transition is quite broad within a narrow range of pressures ($\sim$0.5 GPa), after which, \textit{T$_{c}$} is remarkably enhanced  and the transitions become sharp again for the samples with nominal $x$ = 0.1, 0.3, 0.6. In the case of the $x$ = 0.8 sample, however, the width of the SC transition remains broad at high pressures after the enhancement in \textit{T$_{c}$}; this might be related to the $x$ = 0.8 sample being so close to the solubility limit and possible chemical inhomogeneity of the sample.\cite{Fang} This behavior reveals pressure-induced transitions from a low-\textit{T$_{c}$} superconducting phase (SC1) to a high-\textit{T$_{c}$} superconducting phase (SC2) in each sample.

The abrupt change in \textit{T$_{c}$} under pressure is particularly apparent in the temperature-pressure phase diagram (see Fig. 2), in which \textit{P$_{\rm t}$$^{\rm o}$} is the onset of the phase transition, \textit{P$_{\rm t}$$^{\rm c}$} is the pressure where the phase transition is complete, and \textit{P$_{\rm t}$} is the midpoint between \textit{P$_{\rm t}$$^{\rm o}$} and \textit{P$_{\rm t}$$^{\rm c}$}. The broad superconducting transitions between \textit{P$_{\rm t}$$^{\rm o}$} and \textit{P$_{\rm t}$$^{\rm c}$} indicate the emergence of the SC2 phase; the sample in this pressure range is presumably in a mixture of the SC1 and SC2 phases. The values of \textit{T$_{c}$} obtained in measurements with decreasing pressure (open symbols) are consistent with those measured with increasing pressure (filled symbols), revealing that the pressure-induced phase transitions are fully reversible. Tomita \textit{et al.} recently reported a similar enhancement of \textit{T$_{c}$} at $\sim$0.7 GPa in polycrystalline samples of LaO$_{0.5}$F$_{0.5}$BiS$_{2}$, which was attributed to a structural phase transition from tetragonal to monoclinic due to sliding between two neighboring BiS$_{2}$-layers along the $a$-axis.\cite{TakahiroTomita} Since both LaO$_{0.5}$F$_{0.5}$BiS$_{2}$ and La$_{1-x}$Sm$_{x}$O$_{0.5}$F$_{0.5}$BiS$_{2}$ are characterized by the same crystal structure with space group group \textit{P4/nmm} at ambient pressure and have similar chemical compositions, it seems likely that the pressure-induced enhancement of \textit{T$_{c}$} for La$_{1-x}$Sm$_{x}$O$_{0.5}$F$_{0.5}$BiS$_{2}$ is also associated with a structural phase transition.

\begin{figure}[h]
\centering
\includegraphics[width=8.5cm]{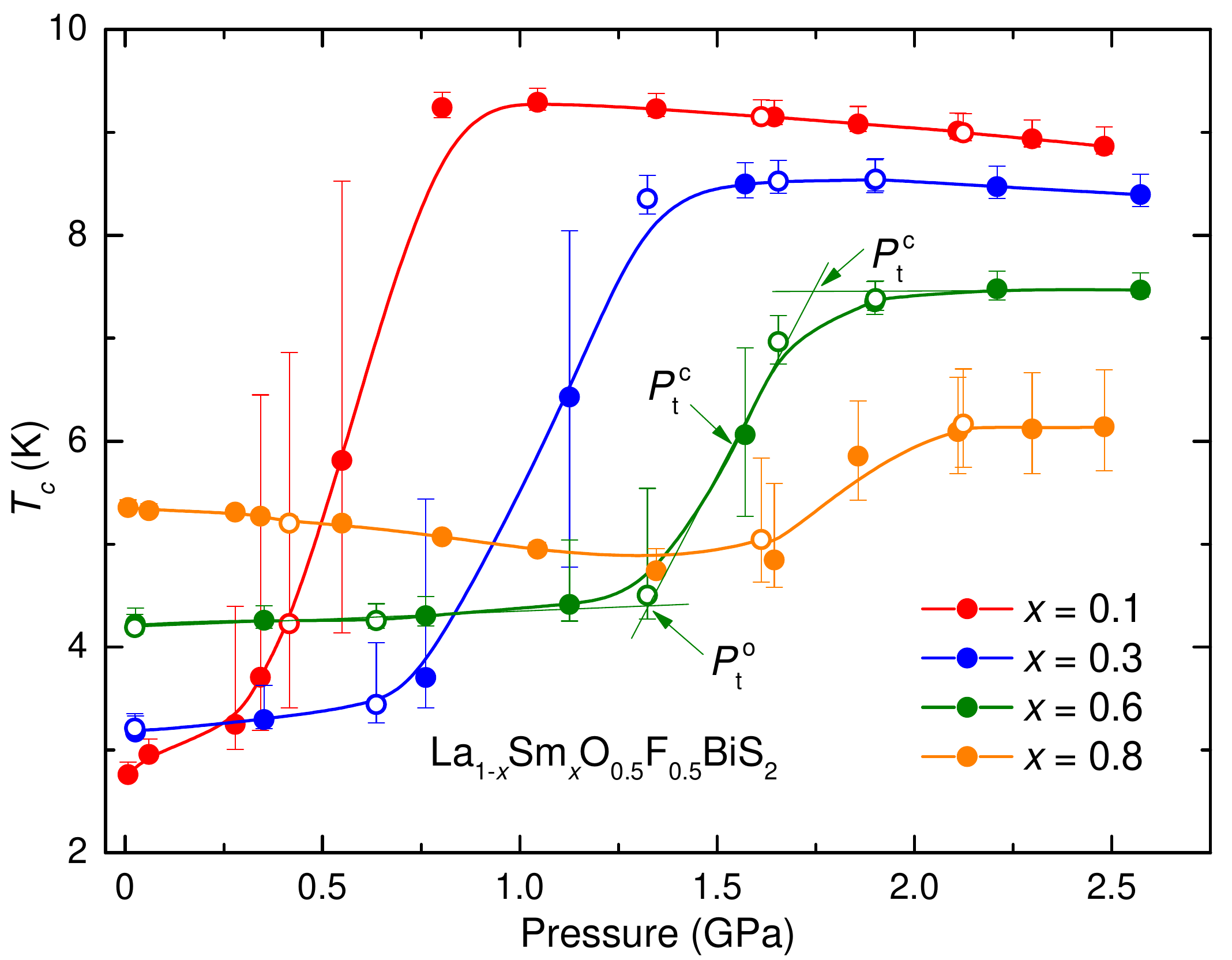}
\caption{(Color online) Superconducting critical temperature \textit{T$_{c}$} vs pressure for La$_{1-x}$Sm$_{x}$O$_{0.5}$F$_{0.5}$BiS$_{2}$ ($x$ = 0.1, 0.3, 0.6, 0.8). Filled and open circles with the same color represent \textit{T$_{c}$} measured with increasing and decreasing pressure, respectively. The width of the superconducting transition is denoted by the vertical bars. \textit{P$_{\rm t}$$^{\rm o}$} and \textit{P$_{\rm t}$$^{\rm c}$} are defined as the pressure of the low-\textit{T$_{c}$} to high-\textit{T$_{c}$} phase transition onsets and completions, respectively;  \textit{P$_{\rm t}$} is the midpoint between \textit{P$_{\rm t}$$^{\rm o}$} and \textit{P$_{\rm t}$$^{\rm c}$}.}
\label{FIG.2.}
\end{figure}

The \textit{T$_{c}$} values of La$_{1-x}$Sm$_{x}$O$_{0.5}$F$_{0.5}$BiS$_{2}$ superconductors at ambient pressure gradually increase from $\sim$2.7 K for $x$ = 0.1 to $\sim$5.4 K for $x$ = 0.8 with increasing Sm substitution up to the solubility limit near $x$ = 0.8.\cite{Fang} However, the pressure dependence of \textit{T$_{c}$} for all of the samples at \textit{P} \textless \textit{P$_{\rm t}$$^{\rm o}$} differ significantly from each other. For the $x$ = 0.1, 0.3, 0.6 samples, \textit{T$_{c}$} is initially enhanced with increasing pressure almost linearly and $d$\textit{T$_{c}$}/$d$\textit{P} decreases with increasing Sm concentration as is shown in Fig. 3 (right axis). Although the value of \textit{T$_{c}$} for $x$ = 0.8 is the highest among the four compositions at ambient pressure, $d$\textit{T$_{c}$}/$d$\textit{P} is negative with a value of roughly -0.45 K/GPa, which results in a decreasing \textit{T$_{c}$} with pressure down to \textit{P$_{\rm t}$$^{\rm o}$}. These results suggest that reduction of the lattice parameter $a$, which is regarded to be essential to the enhancement of \textit{T$_{c}$} for BiS$_{2}$-based compounds at ambient pressure,\cite{Kajitani,Fang,Mizuguchi&Yoshikazu2,Kajitani2} does not always result in an increase of \textit{T$_{c}$}.

\begin{figure}[t]
\centering
\includegraphics[width=8.5cm]{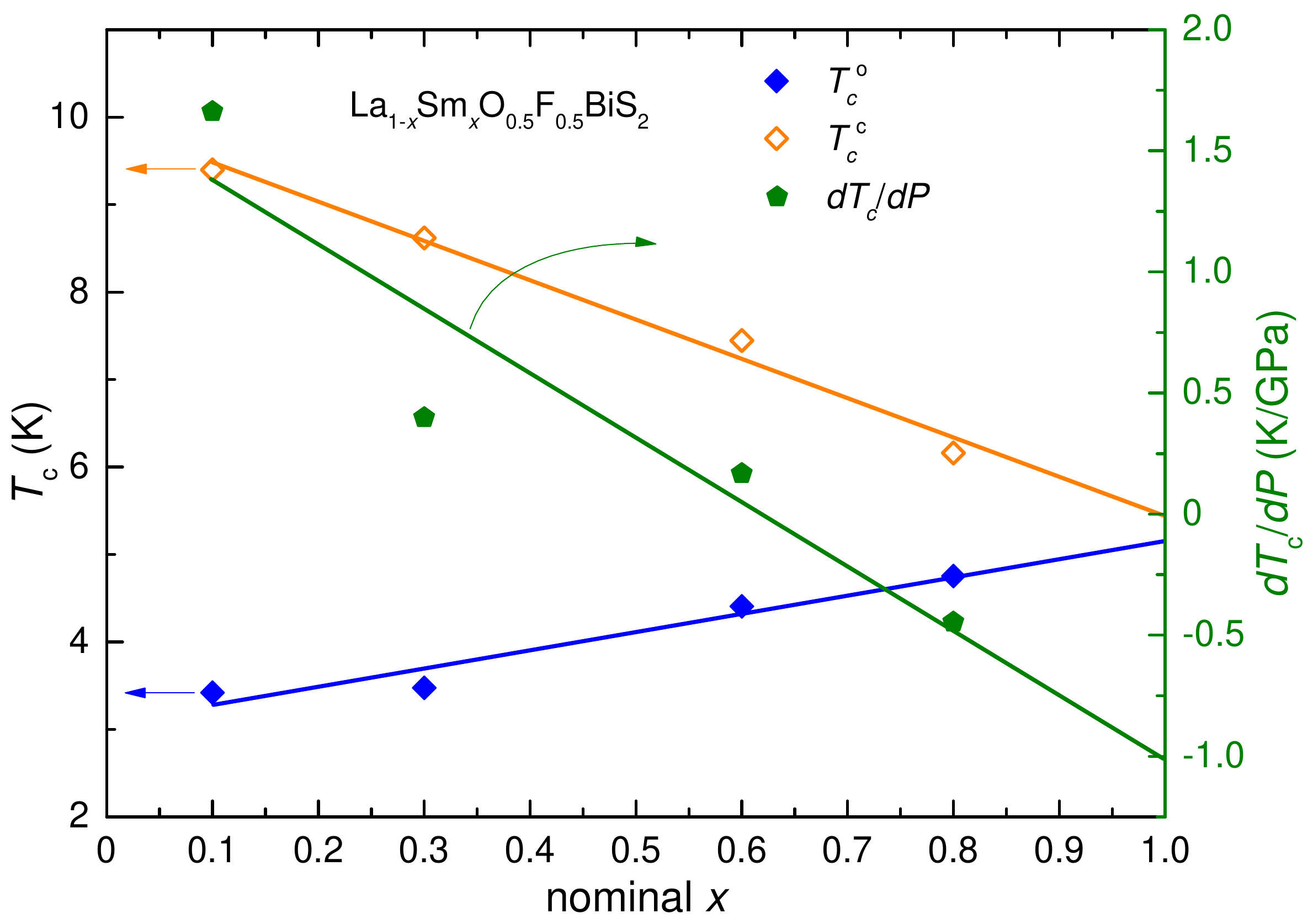}
\caption{(Color online) Sm concentration dependence of \textit{T$_{c}$} at \textit{P$_{\rm t}$$^{\rm o}$} (\textit{T$_{c}$$^{\rm o}$}) and at \textit{P$_{\rm t}$$^{\rm c}$} (\textit{T$_{c}$$^{\rm c}$}) together with the initial rate of change in \textit{T$_{c}$} with pressure ($d$\textit{T$_{c}$}/$d$$P$) for \textit{P} $\textless$ \textit{P$_{\rm t}$$^{\rm o}$}.}
\label{FIG.3.}
\end{figure}

The \textit{T$_{c}$} values at pressures just below and above the transition at \textit{P$_{\rm t}$} from SC1 to SC2 are plotted in Fig. 3 (left axis) and are denoted \textit{T$_{c}$$^{\rm o}$} and \textit{T$_{c}$$^{\rm c}$}, respectively. Although \textit{T$_{c}$$^{\rm o}$} increases with Sm substitution, \textit{T$_{c}$$^{\rm c}$} is suppressed almost linearly from 9.4 K for $x$ = 0.1 to 6.2 K for $x$ = 0.8, resulting in a reduction of the size of the jump in \textit{T$_{c}$} below and above the phase transition ($\Updelta$\textit{T$_{c}$}) with increasing Sm concentration. This phenomenon is very similar to that which was observed for the compounds \textit{Ln}O$_{0.5}$F$_{0.5}$BiS$_{2}$ (\textit{Ln} = La, Ce, Pr, Nd), in which the \textit{T$_{c}$} values above the phase transition decrease with increasing \textit{Ln} atomic number.\cite{Wolowiec1} By linearly extrapolating both \textit{T$_{c}$$^{\rm o}$}  and \textit{T$_{c}$$^{\rm c}$} of La$_{1-x}$Sm$_{x}$O$_{0.5}$F$_{0.5}$BiS$_{2}$ to $x$ = 1, the estimated $\Updelta$\textit{T$_{c}$} of SmO$_{0.5}$F$_{0.5}$BiS$_{2}$ is only $\sim$0.27 K, which is within the uncertainty of our estimate.  If we plot $\Updelta$\textit{T$_{c}$} for \textit{Ln}O$_{0.5}$F$_{0.5}$BiS$_{2}$, La$_{1-x}$Sm$_{x}$O$_{0.5}$F$_{0.5}$BiS$_{2}$, and the estimated $\Updelta$\textit{T$_{c}$} for SmO$_{0.5}$F$_{0.5}$BiS$_{2}$ as a function of lattice parameter $a$, they are located almost on the same line despite the differences in their chemical composition (see Fig. 4(a)). 


\begin{figure}[t]
\centering
\includegraphics[width=8.5cm]{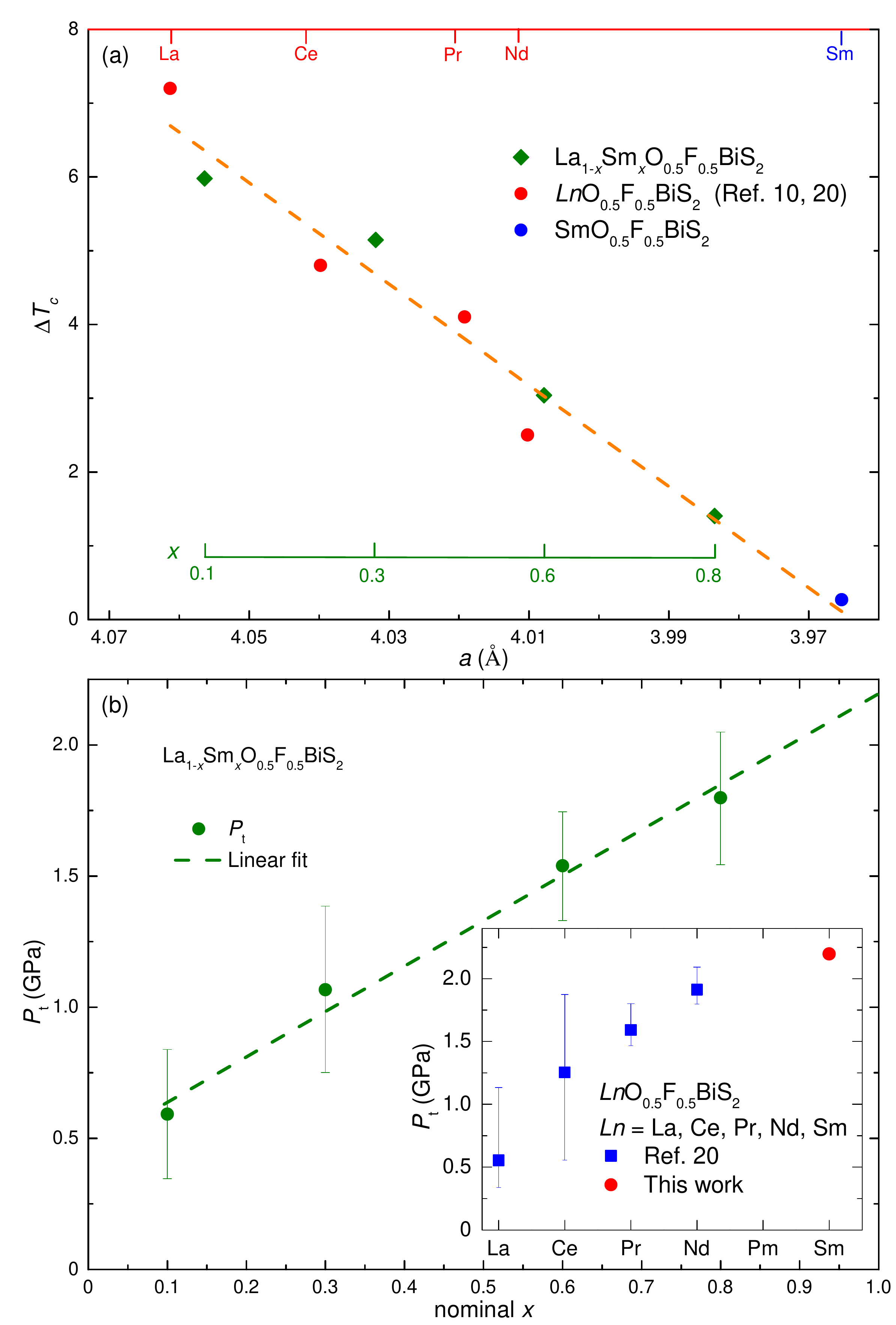}
\caption{(Color online) Dependence of (a) $\Updelta$\textit{T$_{c}$} on lattice parameter $a$ and (b) dependence of \textit{P$_{\rm t}$} on nominal Sm concentration. The inset in panel (b) displays the evolution of \textit{P$_{\rm t}$} as a function of \textit{Ln}. Dashed lines are guides to the eye.}
\label{FIG.4.}
\end{figure}

With Sm substitution for La, \textit{P$_{\rm t}$} increases from 0.59 GPa for $x$ = 0.1 to 1.80 GPa for $x$ = 0.8. The nearly linear evolution of \textit{P$_{\rm t}$} with Sm concentration (see Fig. 4(b)) enables one to estimate a \textit{P$_{\rm t}$}  of $\sim$2.2 GPa for SmO$_{0.5}$F$_{0.5}$BiS$_{2}$, which is consistent with the \textit{P$_{\rm t}$} values of other known \textit{Ln}O$_{0.5}$F$_{0.5}$BiS$_{2}$ compounds as is shown in the inset of Fig. 4(b).\cite{Wolowiec1,Wolowiec2} Figure 5 shows the \textit{T$_{c}$} of La$_{1-x}$Sm$_{x}$O$_{0.5}$F$_{0.5}$BiS$_{2}$ at different pressures and Sm concentrations together with the estimated \textit{T$_{c}$} of SmO$_{0.5}$F$_{0.5}$BiS$_{2}$ under pressure. The \textit{T$_{c}$} at pressures below and above 2.2 GPa of SmO$_{0.5}$F$_{0.5}$BiS$_{2}$ is estimated from a linear extrapolation of the \textit{T$_{c}$} values of the four samples below and above the phase transition, respectively. It can be seen that higher \textit{T$_{c}$} values for the SC1 phase are found at higher Sm concentrations and lower pressures; however, higher \textit{T$_{c}$} values for the SC2 phase are located in the region with lower Sm concentration just above the phase transition. In fact, by further reducing the Sm concentration to $x$ = 0, a higher \textit{T$_{c}$} value of $\sim$10.7 K could be obtained under pressure in LaO$_{0.5}$F$_{0.5}$BiS$_{2}$, which has the largest $a$ lattice parameter among the \textit{Ln}O$_{0.5}$F$_{0.5}$BiS$_{2}$ compounds at ambient pressure.\cite{TakahiroTomita} On the other hand, reducing $a$ enhances \textit{$T_{c}$} at ambient pressure; however, it also reduces $\Updelta$\textit{T$_{c}$} as discussed above. The pressure dependence of \textit{T$_{c}$} below and above 2.2 GPa for  SmO$_{0.5}$F$_{0.5}$BiS$_{2}$  has a similar trend and is almost indistinguishable as is shown in Fig. 5.  Thus, the possibility that SmO$_{0.5}$F$_{0.5}$BiS$_{2}$ has the same phase at low and high pressures cannot be ruled out. Although it was reported that SmO$_{0.5}$F$_{0.5}$BiS$_{2}$ can be synthesized by solid state reaction,\cite{thakur} this result might be helpful in explaining why this parent compound could not be synthesized in other studies using the same method.\cite{Fang,Kajitani}

\begin{figure}[t]
\centering
\includegraphics[width=8.5cm]{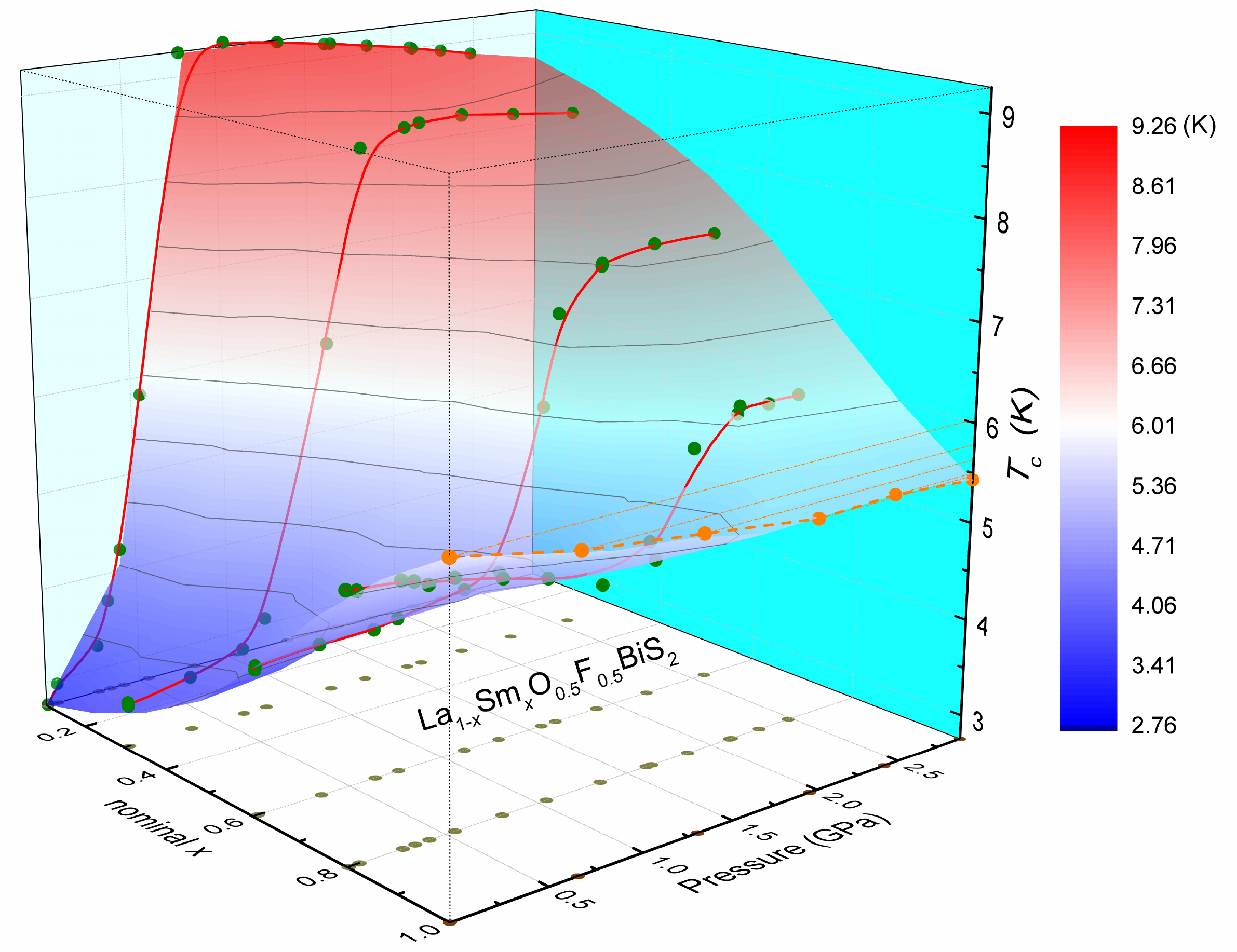}
\caption{(Color online) \textit{T$_{c}$} plotted as a function of pressure and nominal Sm concentration $x$. The \textit{T$_{c}$} of SmO$_{0.5}$F$_{0.5}$BiS$_{2}$ is estimated by linearly extrapolating the \textit{T$_{c}$} of La$_{1-x}$Sm$_{x}$O$_{0.5}$F$_{0.5}$BiS$_{2}$ ($x$ = 0.1, 0.3, 0.6, 0.8) below and above the phase transition near \textit{P$_{\rm t}$}. Filled circles in the $x$-$y$ plane are projections of \textit{T$_{c}$}.}
\label{FIG.5.}
\end{figure}

As is shown in Fig. 6,  $\rho$(\textit{T}) of La$_{1-x}$Sm$_{x}$O$_{0.5}$F$_{0.5}$BiS$_{2}$ increases with decreasing temperature at low pressures in their normal state, indicating semiconducting-like behavior, which is suppressed by applied pressure. However, as the representative $\rho$(\textit{T}) data in the inset of Fig. 6(d) show, a minimum near 262 K ($T$$_{\rm min}$) emerges at pressures above 0.4 GPa. Semiconducting-like behavior is observed in $\rho$(\textit{T}) for $T$ $\textless$ $T$$_{\rm min}$. \textit{T$_{\rm min}$} decreases with increasing pressure and then saturates at $\sim$1.8 GPa, reaching a value of roughly 40 K. Except for the $x$ = 0.1 sample, in which \textit{T$_{\rm min}$} does not saturate until $\sim$2.5 GPa, similar  behavior is observed in the other samples under high pressures. However, the relationship between \textit{T$_{\rm min}$} and Sm concentration is still unclear. The appearance of a metallic-like to semiconducting-like cross over was also found in Eu$_{3}$Bi$_{2}$S$_{4}$F$_{4}$, Sr$_{1-x}$La$_{x}$FBiS$_{2}$ ($x$ $\geqslant$ 0.45), LaOBiS$_{2}$, and ThOBiS$_{2}$ at ambient pressure. In the first example, this behavior was attributed to a self-doping effect arising from Eu intermediate valence, while in the second example, it was ascribed to Anderson localization.\cite{zhai2014anomalous,HideakiSakai,Yazici2} However, further work is required to identify the origin of the minimum in La$_{1-x}$Sm$_{x}$O$_{0.5}$F$_{0.5}$BiS$_{2}$ since the valence of Sm ions might change and sample defects might develop at high pressures.

\begin{figure}[t]
\centering
\includegraphics[width=8.5cm]{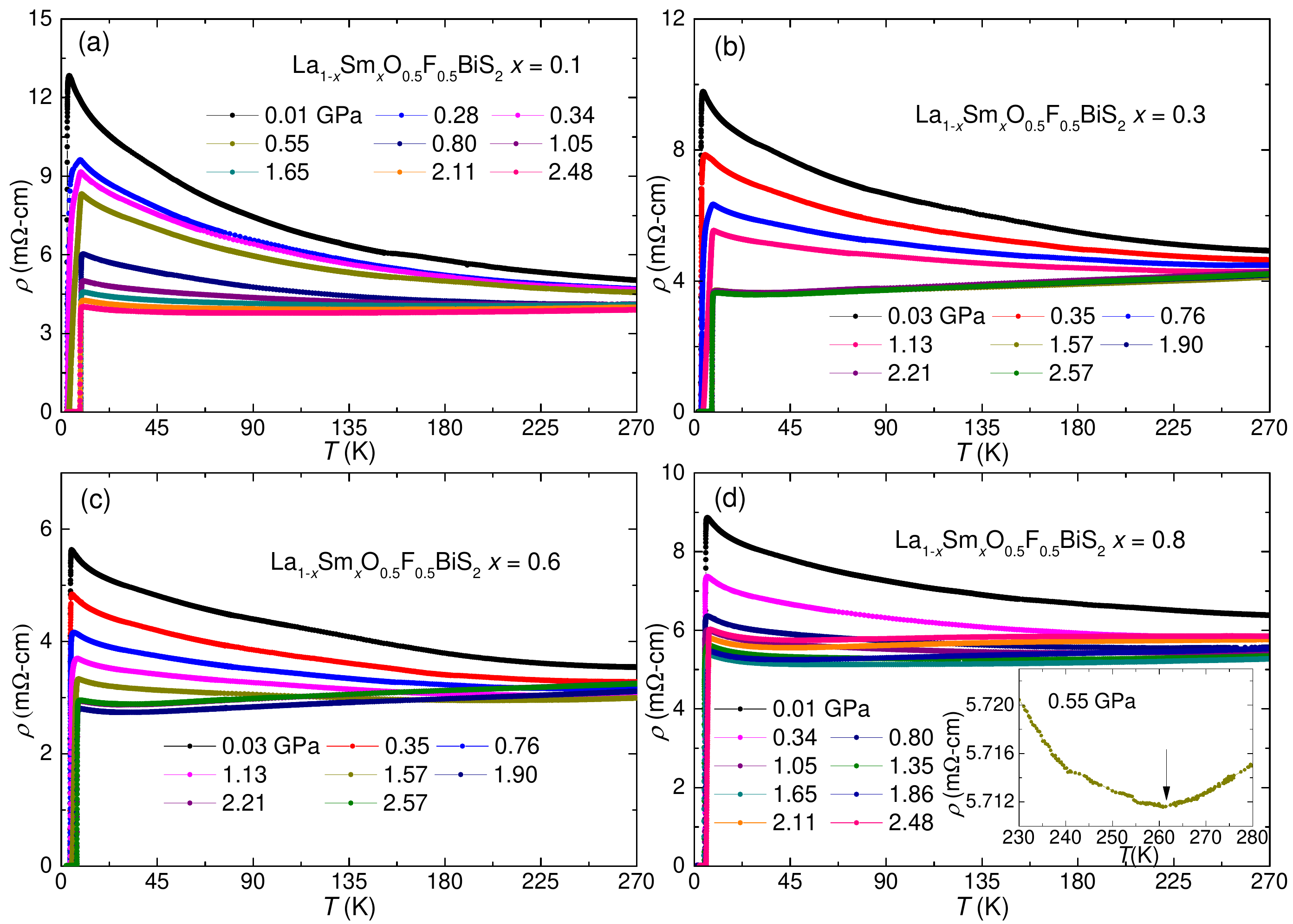}
\caption{(Color online) (a), (b), (c), (d) Electrical resistivity of La$_{1-x}$Sm$_{x}$O$_{0.5}$F$_{0.5}$BiS$_{2}$ with $x$ = 0.1, 0.3, 0.6, and 0.8 under applied pressure from 270 K to $\sim$1.2 K, respectively. The inset of panel (d) is a representative $\rho$($T$) curve in which a minimum (indicated by the arrow) is observed in the normal state upon cooling.}
\label{FIG.6.}
\end{figure}

\begin{figure}[h]
\centering
\includegraphics[width=8.5cm]{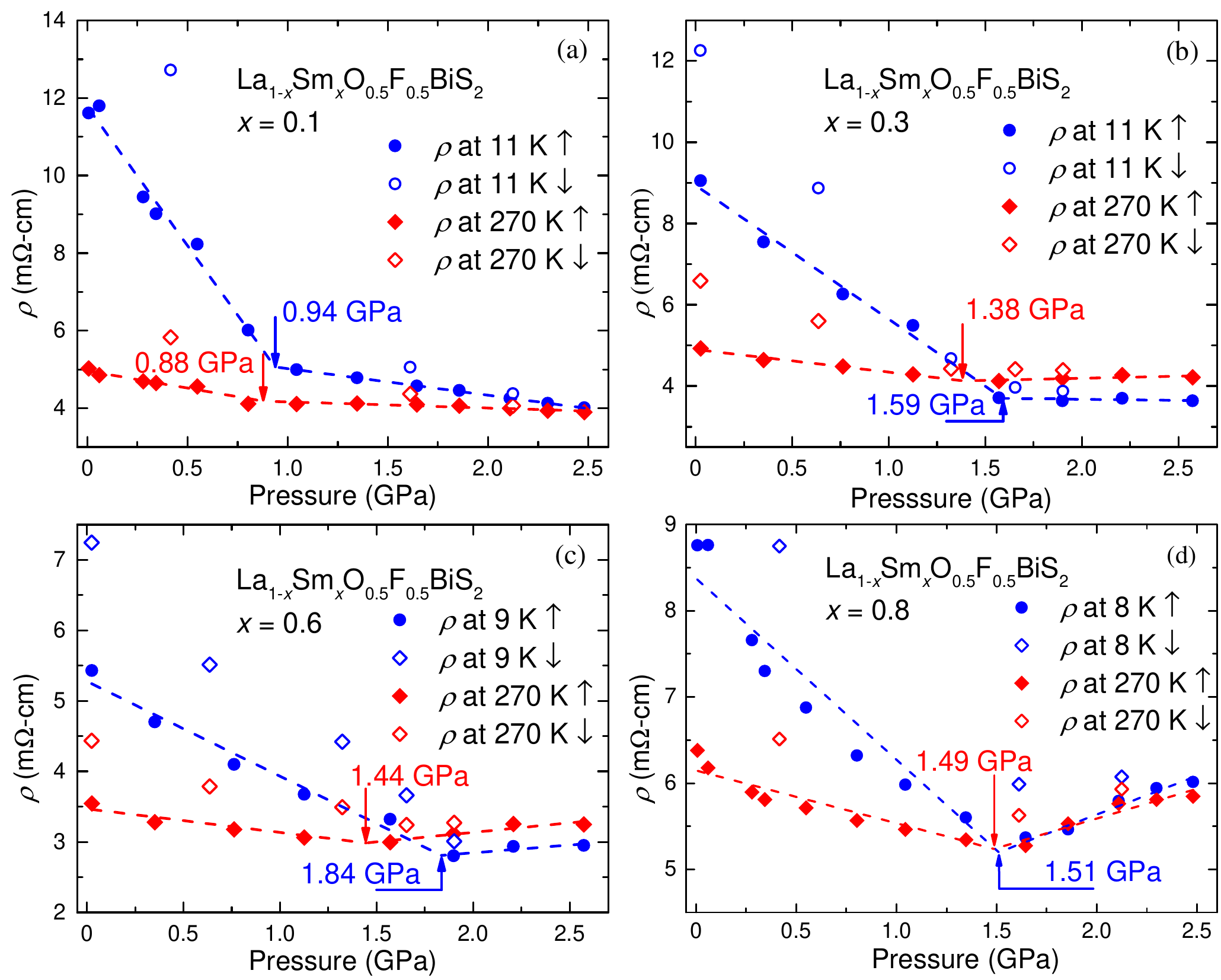}
\caption{(Color online) (a), (b), (c), (d) Electrical resistivity of La$_{1-x}$Sm$_{x}$O$_{0.5}$F$_{0.5}$BiS$_{2}$ with $x$ = 0.1, 0.3, 0.6, and 0.8, respectively, at a temperature just above \textit{T$_{c}$} in the normal state and at 270 K. The filled circles represent data collected with increasing pressure and the open circles are the data taken upon releasing the pressure. The dashed lines, which are guides to the eye, reflect the slopes of the pressure dependence of the electrical resistivity. Arrows indicate the pressure at which the slope changes.}
\label{FIG.7.}
\end{figure}

Figure 7 shows $\rho$ values at 270 K ($\rho$$_{\rm 270K}$) and at certain low temperatures ($\rho$$_{\rm low}$), explicitly indicated in the figure, just higher than \textit{T$_{c}$} for each sample. Both $\rho$$_{\rm 270K}$ and $\rho$$_{\rm low}$ first decrease with increasing pressure almost linearly. However, the slope of the $\rho$($P$) curve changes at the pressure indicated by the arrows in Fig. 7, resulting in a kink for each $\rho$($P$) curve. For the $x$ = 0.6 and 0.8 samples, a positive pressure coefficient of resistivity is observed at the pressures above the kink. Although the kink is located at pressures near \textit{P$_{\rm t}$}, it should be noted that the kink does not necessarily coincide with the appearance of the SC2 phase, since both the phase transition between SC1 to SC2 and the appearance of metallic-like behavior may contribute to the normal-state electrical resistivity of the samples.

Since the electrical transport behavior of single crystals of La$_{1-x}$Sm$_{x}$O$_{0.5}$F$_{0.5}$BiS$_{2}$ has not yet been reported, it is still difficult to determine whether the observed semiconducting-like behavior is related to poor intergain contact or whether it is an intrinsic property. If we assume that the semiconducting-like behavior is intrinsic, the energy gap can be estimated by using the simple activation-type relation:

  \begin{equation}
  \rho(T)=\rho_{0}e^{\Updelta/2k_{B}T},
  \end{equation}
where $\rho$$_{0}$ is a constant, $k$$_{B}$ is the Boltzmann constant, and $\Updelta$ is the energy gap.\cite{HisashiKotegawa,Inho} Because the relationship between 1/$T$ and ln$\rho$ is not linear for the whole temperature range in the normal state, the $\rho$(\textit{T}) data were fitted using Eq. (1) in two different temperature ranges, 270-160 K and 22-10 K, similar to the analysis in Refs. \onlinecite{Wolowiec1, Wolowiec2, HisashiKotegawa,Inho}. The energy gaps $\Updelta$$_{1}$/$k$$_{B}$ and $\Updelta$$_{2}$/$k$$_{B}$, which are extracted from the high temperature and low temperature ranges, respectively, are plotted in Fig. 8. The gap $\Updelta$$_{1}$/$k$$_{B}$ is not shown above the pressure at which metallic-like conduction is observed. It can be seen that both $\Updelta$$_{1}$/$k$$_{B}$ and $\Updelta$$_{2}$/$k$$_{B}$ are suppressed rapidly with increasing pressure; however, the value of $\Updelta$$_{1}$/$k$$_{B}$ does not change significantly above \textit{P$_{\rm t}$$^{\rm c}$} with increasing pressure for all of the samples. Hence, the SC2 phase probably has a different band structure which is less sensitive to pressure.

\begin{figure}[t]
\centering
\includegraphics[width=8.5cm]{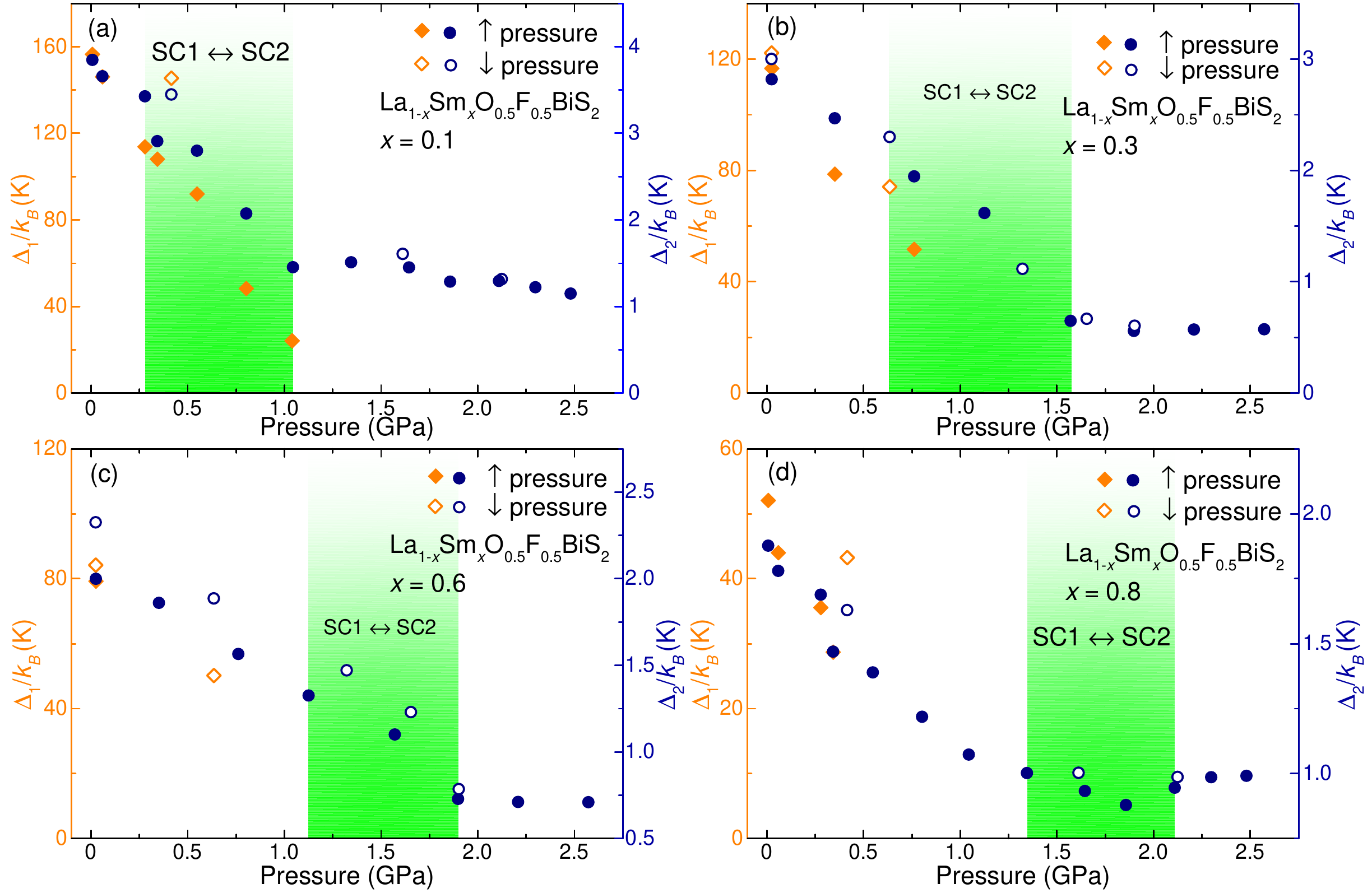}
\caption{(Color online) (a), (b), (c), (d) Evolution of the energy gaps $\Updelta$$_{1}$/$k$$_{B}$ and $\Updelta$$_{2}$/$k$$_{B}$ plotted as a function of pressure for $x$ = 0.1, 0.3, 0.6, and 0.8, respectively. Filled and open symbols represent values that were obtained with increasing and decreasing pressure, respectively. Pressure ranges in which the low-\textit{T$_{c}$} to high-\textit{T$_{c}$} phase transition is observed and in which the normal-state electrical resistivity shows metallic-like behavior above 40 K are also indicated.}
\label{FIG.8.}
\end{figure}

Although \textit{T$_{c}$} is reversible (i.e., there is no difference in values measured for increasing and decreasing pressure), values of $\Updelta$$_{1}$/$k$$_{B}$ and $\Updelta$$_{2}$/$k$$_{B}$ are higher for decreasing pressure than increasing pressure, indicating an enhancement of semiconducting-like behavior for all of the samples. A similar phenomenon was also observed in LaO$_{0.5}$F$_{0.5}$BiS$_{2}$ and CeO$_{0.5}$F$_{0.5}$BiS$_{2}$.\cite{Wolowiec2} This behavior is associated with the pressure dependence of the electrical resistivity in which values of $\rho$ obtained at similar pressures in measurements in which the pressure cell was being unloaded are significantly higher than those measured upon loading as the representative data show in Fig. 7. However, irreversible defects and disorder as well as a reduction of the geometric factors might develop in the samples under high pressures, which could contribute to a reduction in the conductivity of the samples. Since the energy gap $\Updelta$$_{1}$/$k$$_{B}$ and $\Updelta$$_{2}$/$k$$_{B}$ are derived from the electronic structure, there is no obvious reason why these fundamental quantities should show hysteretic behavior as a function of pressure.

\section{SUMMARY}

Electrical resistivity measurements on polycrystalline samples of the BiS$_{2}$-based superconductors La$_{1-x}$Sm$_{x}$O$_{0.5}$F$_{0.5}$BiS$_{2}$ ($x$ = 0.1, 0.3, 0.6, 0.8) were performed from 2 K to room temperature under applied pressures. In the normal state, semiconducting-like behavior is suppressed with increasing pressure. A reversible low-\textit{T$_{c}$} to high-\textit{T$_{c}$} superconductor phase transition was observed in all of the samples at a pressure that is proportional to the Sm concentration. With increasing Sm concentration, $\Updelta$\textit{T$_{c}$} is suppressed and a larger pressure is necessary to induce the transition from the SC1 to the SC2 phase. It is also found that an optimal \textit{T$_{c}$}  could be tuned by decreasing the $a$ lattice parameter in the SC1 phase at ambient pressure or by increasing $a$ in the SC2 phase under pressure. These results indicate that the high-pressure behavior of Sm-substituted  LaO$_{0.5}$F$_{0.5}$BiS$_{2}$ is largely determined by Sm concentration or the lattice parameter $a$ at ambient pressure. Therefore, the evolution of \textit{T$_{c}$} under pressure for the parent compound SmO$_{0.5}$F$_{0.5}$BiS$_{2}$ can be estimated, and we find that the SC1 and SC2 phases exhibit almost indistinguishable \textit{T$_{c}$} values.

\begin{acknowledgements}
Electrical resistivity measurements under applied pressure were supported by the National Nuclear Security Administration under the Stewardship Science Academic Alliance Program through the US Department of Energy (DOE) under Grant No. DE-NA0001841. Sample synthesis and characterization at ambient pressure were supported by the U. S. Department of Energy, Office of Basic Energy Sciences, Division of Materials Sciences and Engineering under Grant No. DE-FG02-04-ER46105. Helpful discussions with I. Jeon and C. T. Wolowiec are gratefully acknowledged.
\end{acknowledgements}

\bibliography{Manuscript}
\clearpage

\end{document}